\documentclass[12pt,a4paper]{article}
\usepackage[pdftex,colorlinks=true]{hyperref}
\usepackage{amsmath, amssymb, natbib, setspace, bm, graphicx, subfig, booktabs, todonotes, tikz}
\usepackage[neverdecrease]{paralist}
\usepackage[hmargin=1in,vmargin=1in]{geometry}

\usepackage{enumitem}
\setlist{noitemsep, topsep=0pt, parsep=0pt, partopsep=0pt, leftmargin=1.2cm}

\definecolor{Rcolor}{RGB}{150,160,190}

\newcommand{\Rx}{\fontsize{10pt}{12pt}\selectfont
\raisebox{.3em}{\hspace{1.2em}%
\llap{\resizebox{1.09em}{.5em}{\color{black}$\bigcirc$}}%
\llap{\resizebox{1.199em}{.55em}{\color{darkgray}$\bigcirc$}}%
\llap{\resizebox{1.19em}{.52em}{\color{gray!50}$\bigcirc$}}%
\llap{\resizebox{1.1em}{.5em}{\color{gray}$\bigcirc$}}%
\llap{\resizebox{1.25em}{.55em}{\color{gray}$\bigcirc$}}%
}%
\hspace{-.85em}%
\textbf{%
\textcolor{black}{\textsf{R}}%
\hspace{-.025em}\raisebox{.01em}{\llap{\textcolor{Rcolor}{\textsf{R}}}}%
}}%

\newbox\rbox
\savebox\rbox{\scalebox{0.1}{\Rx}}

\makeatletter
\def\R{\scalebox{\f@size}{\usebox\rbox}\xspace}
\makeatletter

\graphicspath{{plots/}}

\hypersetup{citecolor=blue,linkcolor=blue,urlcolor=blue}
\newsavebox\CBox
\def\textBF#1{\sbox\CBox{#1}\resizebox{\wd\CBox}{\ht\CBox}{\textbf{#1}}}
\begin{document}

\begin{center}
\large On the selection of optimal Box-Cox transformation parameter for modeling and forecasting age-specific fertility
\end{center}

\begin{center}
Han Lin Shang\footnote{Postal address: Research School of Finance, Actuarial Studies and Applied Statistics, Building 26C, Australian National University, Canberra 0200, Australia; Telephone: +61 (2) 6125 0535; Email: hanlin.shang@anu.edu.au}\\
Australian National University
\end{center}

\hspace{.5in}

\begin{abstract}

The Box-Cox transformation can sometimes yield noticeable improvements in model simplicity, variance homogeneity and precision of estimation, such as in modeling and forecasting age-specific fertility. Despite its importance, there have been only few studies focusing on the optimal selection of Box-Cox transformation parameter in demographic forecasting. A simple method is proposed for selecting the optimal Box-Cox transformation parameter, along with an algorithm based on an in-sample forecast error measure. Illustrated by Australian age-specific fertility, the out-of-sample accuracy of a forecasting method can be improved with the selected Box-Cox transformation parameter. Furthermore, the log transformation is not adequate for modeling and forecasting age-specific fertility. It is recommended to embed the selection of Box-Cox transformation parameter into statistical analysis of age-specific demographic data, in order to fully capture forecast uncertainties.

\vspace{.3in}

\end{abstract}

\newpage
\section{Introduction}

In the demographic literature, forecasting methods for age-specific fertility can be generally grouped into parametric, semi-parametric and nonparametric models. Parametric models used in forecasting include the beta, gamma, double exponential and Hadwiger functions \citep{KMR93, TBL+89, Congdon90, Congdon93, KP00}, while semi-parametric models include the Coale-Trussell and Relational Gompertz models \citep{CT74, Brass81, Murphy82, Booth84, ZWM+00}. The use of these models is variously limited by  parameter uninterpretability, over-parameterization and the need for vector autoregression; structural change also limits their utility, especially where vector autoregression is involved \citep{Booth06}. To address this problem, nonparametric methods use a dimension-reduction technique, such as principal components analysis, to linearly transform age-specific fertility rates to extract a series of time-varying indexes to be forecast \citep[see][]{BB87,Bell92,Lee93, HU07}.  

The Box-Cox transformation can sometimes yield noticeable improvements in model simplicity, variance homogeneity and precision of estimation. Despite the rapid development in demographic forecasting models, there have been only few studies focusing on the optimal selection of the Box-Cox transformation parameter, with an noticeable exception of \cite{HB08}. As noted in early work by \citet{BC64} and \cite{Box88}, the careful selection of a data transformation is often treated as a prerequisite before any serious modeling takes place. 

An example of data transformation is the log transformation for modeling and forecasting age-specific mortality. Such a transformation allows researchers to visualize and model patterns associated with the so-called ``accident bump" and to exploit near-linearities in the log mortality rates for the ages 40 to 80 years. The log transformation is a special case of the Box-Cox transformation, which can be defined as
\[ z_{t,i} = \left\{ \begin{array}{ll}
         \frac{1}{\lambda}\left[(f_{t,i})^{\lambda}-1\right] & \mbox{if $\lambda \neq 0$}\\
        \ln (f_{t,i}) & \mbox{if $\lambda = 0$}\end{array} \right.\]
where $f_{t,i}>0$ denotes the observed age-specific data at age $i$ in year $t$, whereas $z_{t,i}$ denotes the transformed data, and $\lambda$ is the transformation parameter. For instance, when $\lambda=1$, the transformation is essentially the identity, and the logarithm when $\lambda=0$. In this work, we restrict it to lie in the unit interval \citep[see also][]{HB08}. 

We propose a simple and instructive way of selecting the optimal Box-Cox transformation parameter based on an in-sample forecast error measure, and to demonstrate this idea in the context of modeling and forecasting age-specific fertility. The effect of the Box-Cox transformation on fertility is mainly manifested by a different shape of age profile. With the optimal transformation parameter, the age profile of the transformed data may reveal age patterns that are not obvious in the raw data.

This paper is organized as follows. In Section~\ref{sec:2}, we present the Australian age-specific fertility from 1921 to 2006. In Section~\ref{sec:3}, we present the methodology and optimization algorithm. Results are collated in Section~\ref{sec:4}. Section~\ref{sec:5} concludes, along with some thoughts on how the method developed here might be further extended.

\section{Data and design}\label{sec:2}

\subsection{Data set}

\begin{figure}[!ht]
\centering
\subfloat[Rainbow plot of fertility rates]
{\includegraphics[width=7.7cm]{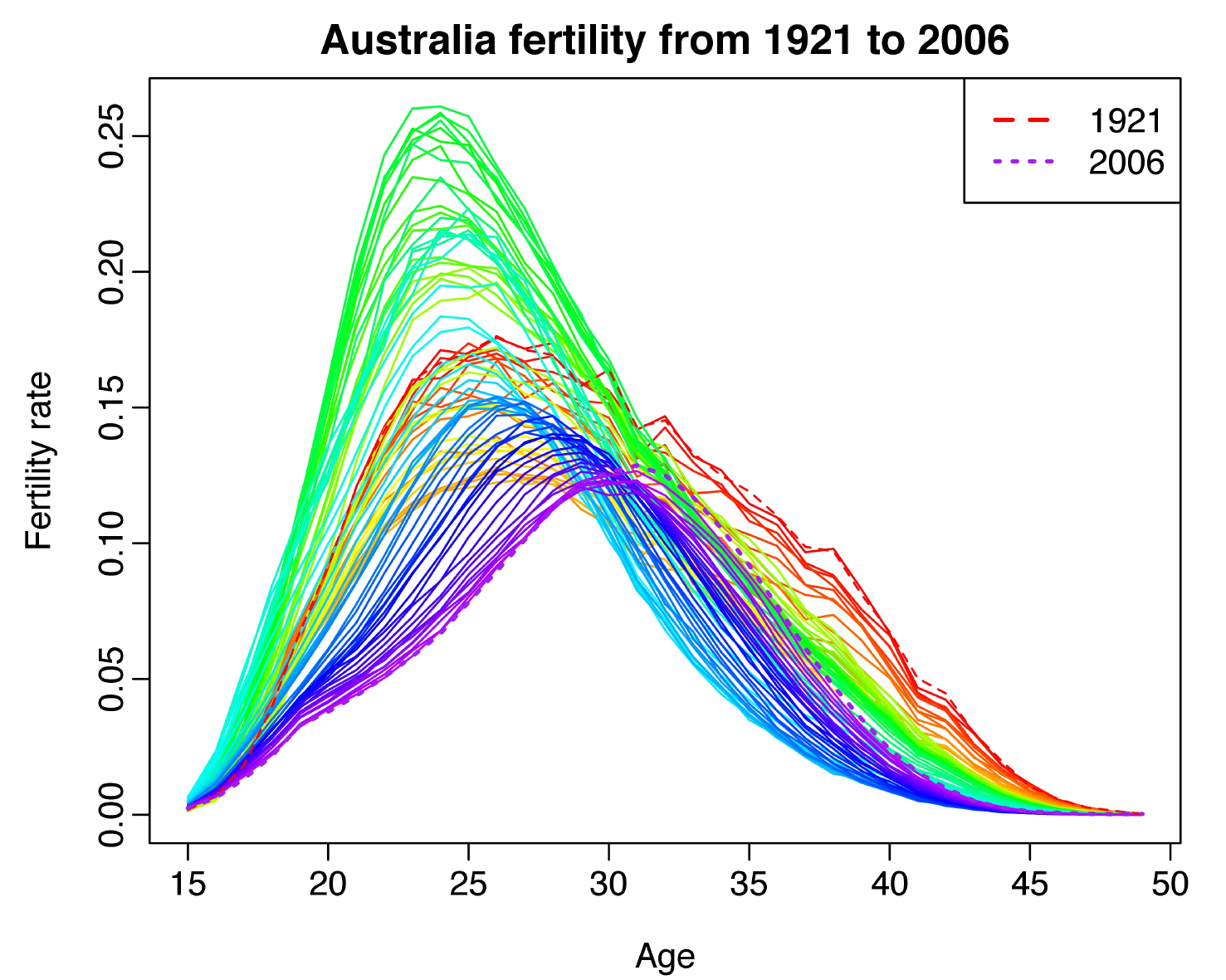}\label{fig:1a}}
\quad
\subfloat[Filled contour plot of fertility rates]
{\includegraphics[width=7.7cm]{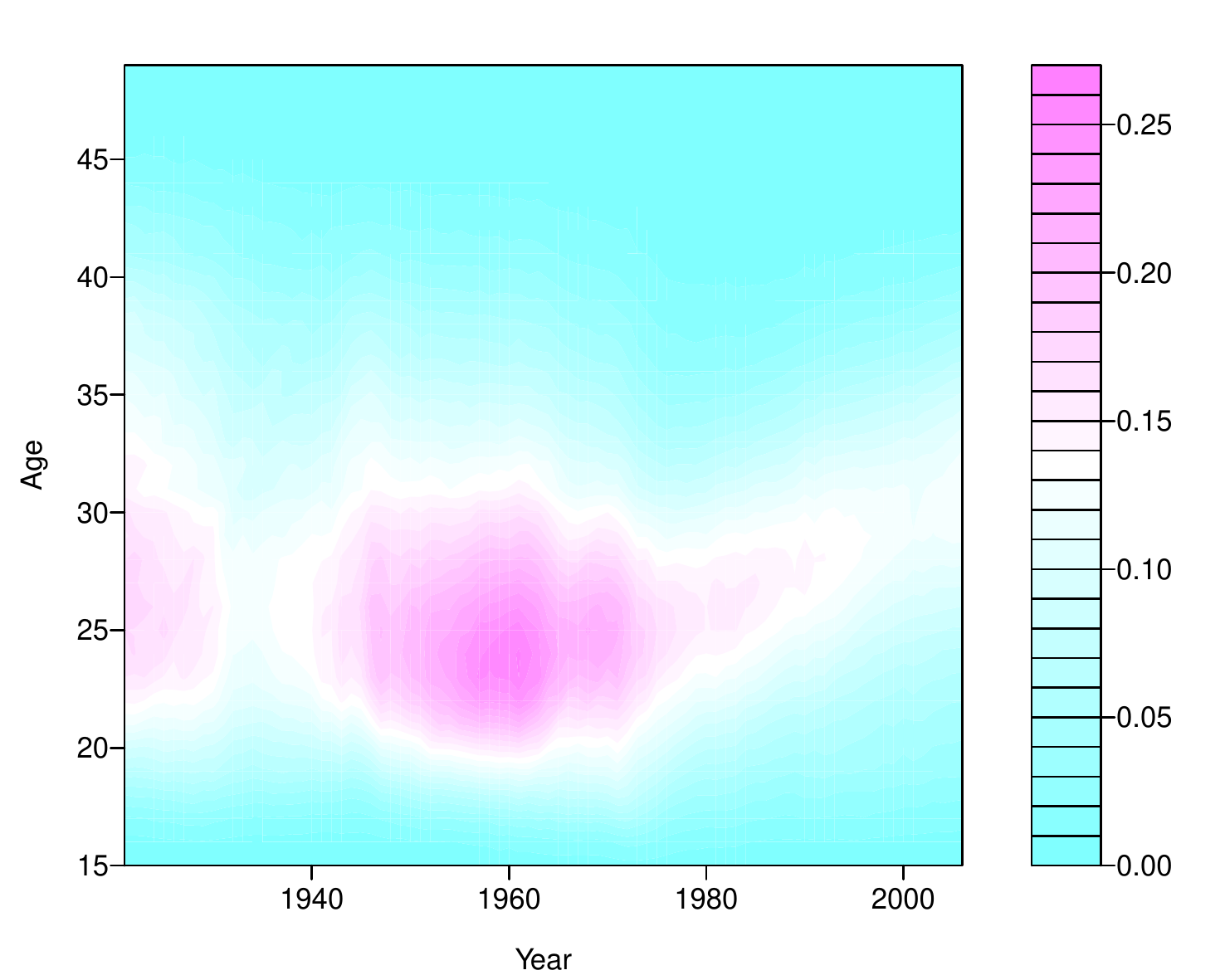}\label{fig:1b}}
\caption{\onehalfspacing Observed age-specific fertility rates for Australia from 1921 to 2006. In Figure (a), the dashed line represents the data in 1921, while the dotted line represents the data in 2006.
Source: Australian Bureau of Statistics (Cat. No. 3105.0.65.001, Table 38).}\label{fig:1}
\end{figure}

We consider annual Australian age-specific fertility rates from 1921 to 2006. The data set has been obtained from the Australian Bureau of Statistics (Cat. No. 3105.0.65.001, Table 38), and is also available in the \textit{rainbow} package \citep{SH13} in  {\Rx{}} \citep{Team13}. The data consist of annual fertility rates by single-year age of mother aged from 15 to 49 years. A graphical data display is given in Figure~\ref{fig:1}. From the rainbow plot in Figure~\ref{fig:1a}, we see the phenomenon of fertility postponement in the most recent years. From the contour plot in Figure~\ref{fig:1b}, we see the increases in fertility between ages 20 and 30 from 1940 to 1980, this reflects the baby boom period.

\begin{figure}
\centering
\includegraphics[width=7.52cm]{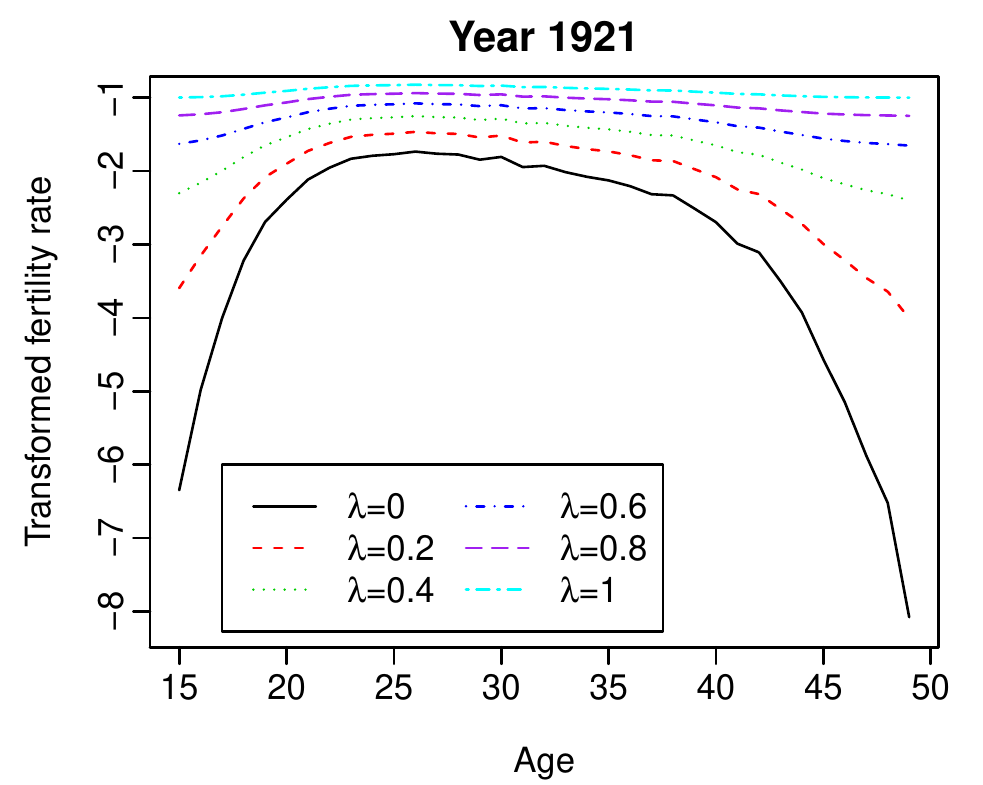}
\quad
\includegraphics[width=7.52cm]{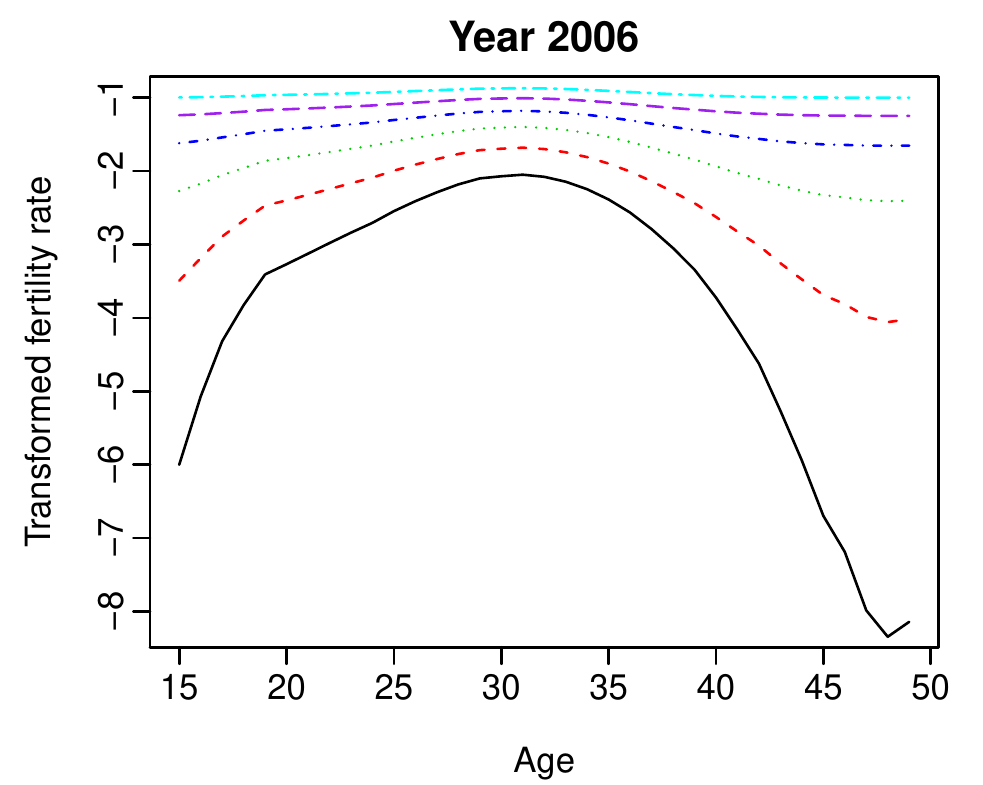}
\caption{Box-Cox transformed fertility rates with different values of $\lambda$ in years 1921 and 2006, as two examples.}\label{fig:BC_example}
\end{figure}

As a demonstration, Figure~\ref{fig:BC_example} presents the Box-Cox transformed fertility rates for years 1921 and 2006. With different values of $\lambda$, the age profiles change accordingly. The goal is to select the optimal $\lambda$ that improves model estimation and prediction accuracy for a chosen model.

\subsection{Study design}

Since the optimal Box-Cox transformation parameter is selected based on an in-sample forecast error measure, we divide the data into a training sample, a validation sample and a testing sample. Customarily, the testing sample consists of the last 20\% of the data, which are used to examine the out-of-sample forecast accuracy with the selected Box-Cox transformation parameter. The validation sample, which has the same number of data as the testing sample, is used to select the optimal Box-Cox transformation parameter. As in the case of Australian fertility rates, the training sample is from 1921 to 1972, the validation sample is from 1973 to 1989, and the testing sample is from 1990 to 2006.

There are various ways to measure the forecast accuracy. Following the early work by \cite*{SBH11}, we use mean absolute forecast error (MAFE) for measuring point forecast accuracy. This is given by
\begin{align*}
  \text{MAFE}_h &= \frac{1}{(18-h)\times 35} \sum^{18-h}_{s=1}\sum^{49}_{i=15}\left|f_{i,s}-\widehat{f}_{i,s}\right|, \qquad h=1,\dots,17. \label{eq:mafe}
\end{align*}
The MAFE is the average of absolute error across ages and years in the forecasting period; it measures forecast precision regardless of sign and is not sensitive to large relative errors of small rates. Since the back-transformed forecasts are median forecasts on the original scale, this makes them suitable for evaluation using MAFE. 

In order to evaluate the interval forecast accuracy, we utilize the interval score of \cite{GR07} and \cite{GK14}. For each year in the forecasting period, the one-step-ahead to 17-step-ahead prediction intervals were calculated at the $(1-\alpha)\times 100\%$ nominal coverage probability. We consider the common case of the symmetric $(1-\alpha)\times 100\%$ prediction interval, with lower and upper bounds that are predictive quantiles at $\alpha/2$ and $1-\alpha/2$, denoted by $l$ and $u$. As defined by \citet{GR07}, a scoring rule is given by the associated interval forecast $S_{\alpha}(l,u;i)$. This can be expressed as
\begin{equation}
  S_{\alpha}(l,u;i) = (u-l)+\frac{2}{\alpha}\left[(l-i)I\{i<l\}+(i-u)I\{i>u\}\right],\label{eq:int_score}
\end{equation}
where $I\{\cdot\}$ represents the binary indicator function which takes the value of 1 when the condition is met, and $\alpha$ denotes the level of significance. In this paper, $\alpha=0.2$ since we construct 80\% prediction interval. The optimal score is achieved when $i$ lies between $l$ and $u$, and the distance between $l$ and $u$ is minimal. The interval score can be interpreted as: a forecaster is rewarded for narrow width of a prediction interval, if and only if the true observation lies within the prediction interval. The smaller the interval score is, the better the method is for producing interval forecasts.

For different ages and years in the forecasting period, the averaged interval score is defined by
\begin{equation*}
S_{\alpha}^{\text{ave}}(l,u;i) = \frac{1}{(18-h)\times 35}\sum_{s=1}^{18-h}\sum^{49}_{i=15}S_{\alpha,s}(l,u;i), \qquad h=1,\dots, 17.
\end{equation*}

\section{Methodology}\label{sec:3}

Many methods have been proposed for modeling age-specific fertility \citep[see][for reviews]{Booth06}.  To demonstrate our main idea, we model the observed period age-specific fertility, using the well-known Lee-Carter model \citep{LC92}. Instead of retaining only the first component, we retain more than one principal component \citep*[see also][]{CBD06}. The modified Lee-Carter model can be defined by
\begin{equation*}
  z_{t,i}=\mu_i + \sum^K_{k=1}\beta_{t,k}\phi_{k,i}+\varepsilon_{t,i},\qquad 1\leq t \leq n, \quad 1\leq i \leq p,
\end{equation*}
where $n$ denotes the last year in the training sample and $p$ denotes the last age, $\mu_i$ represents the mean estimated by $\frac{1}{n}\sum^n_{t=1}z_{t,i}$, $\{\beta_{1,k},\dots,\beta_{n,k}\}$ represents the $k$th estimated principal component scores, $\{\phi_{k,1},\dots,\phi_{k,p}\}$ represents the $k$th estimated principal component which can be obtained from singular value decomposition applied to the training sample, $\varepsilon_{t,i}$ represents the independent and identically distributed Gaussian white noise, and $K$ represents the number of retained principal components and the value of $K$ can be determined by a ratio-based estimator \citep[see][for details]{LYB11}.

\subsection{Point and interval forecasts}\label{sec:3.1}

Conditional on the estimated mean $\widehat{\mu}_i$ and the estimated principal components $\big(\widehat{\phi}_{1,i},\dots,\widehat{\phi}_{K,i}\big)$, the point forecasts are given by
\begin{align*}
  \widehat{z}_{n+h|n,i} &= \widehat{\mu}_i+\sum^K_{k=1}\widehat{\beta}_{n+h|n,k}\widehat{\phi}_{k,i},
\end{align*}
where $\widehat{\beta}_{n+h|n,k}$ represents the $h$-step-ahead point forecast of the $k$th principal component scores. These forecasts can be obtained from applying a univariate time-series model, such as an autoregressive integrated moving average (ARIMA) model. We use the \verb auto.arima \ algorithm of \cite{HK08} to select the optimal orders of an ARIMA model on the basis of an information criterion, such as the corrected Akaike information criterion \citep{HT89} considered in this paper.

Similarly, conditional on the estimated mean and estimated principal components, the total variance can be approximated by
\begin{align*}
 \text{Var}[\widehat{z}_{n+h|n,i}]&\approx \sum^K_{k=1}\widehat{u}_{n+h|n,k}\widehat{\phi}_{k,i}^2+\widehat{v}_{n+h,i},
\end{align*}
where $\widehat{u}_{n+h|n,k}$ denotes the estimated variance of the sample principal component scores;  $\widehat{\phi}_{k,i}^2$ denotes the square of the fixed principal components; and $\widehat{v}_{n+h,i}$ denotes the estimated variance of the model residual \citep*[see also][]{SBH11}. The 80\% prediction interval of the transformed data can be obtained based on the estimated total variance and a normality assumption.

Having obtained the point and interval forecasts for the transformed data, we then back-transform these forecasts to the original scale through inverse Box-Cox transformation. This can be expressed as
\[ \widehat{f}_{n+h|n,i} = \left\{ \begin{array}{ll}
         \left(\lambda \widehat{z}_{n+h|n,i} +1\right)^{\frac{1}{\lambda}} & \mbox{if $\lambda \neq 0$}\\
        \exp (\widehat{z}_{n+h|n,i}) & \mbox{if $\lambda = 0$}\end{array} \right.\]

\subsection{Application to age-specific fertility}

\begin{figure}[!ht]
\centering
\includegraphics[width=\textwidth]{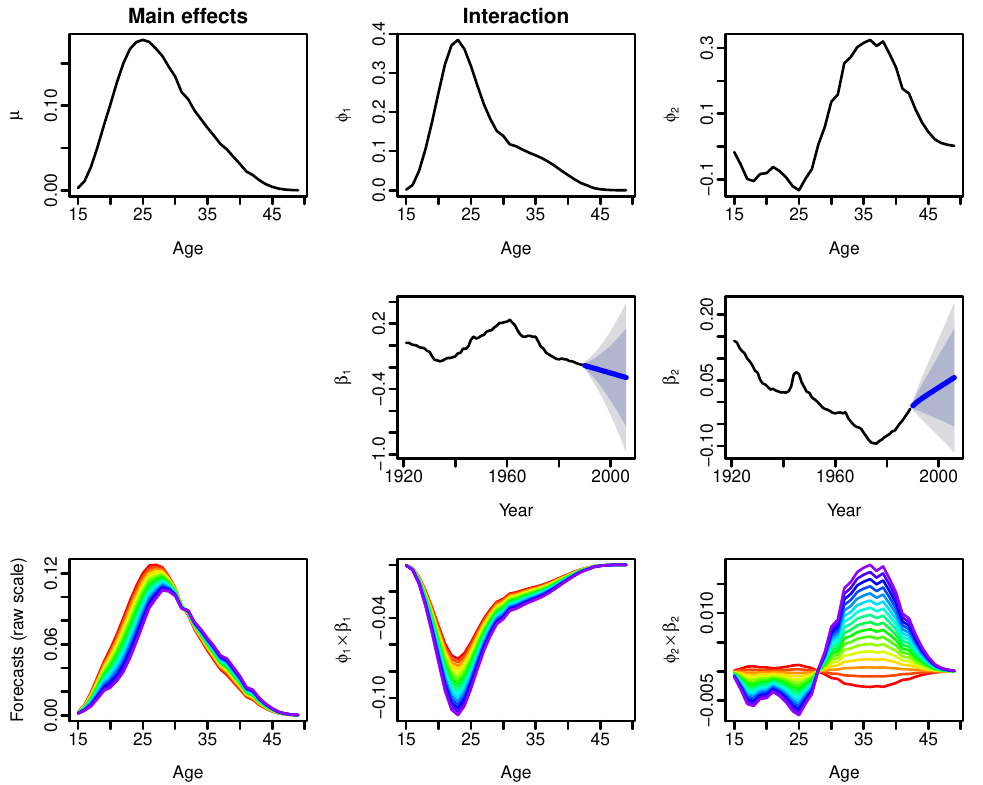}
\caption{Principal component decomposition for the Australian fertility data from 1921 to 1989, from which the forecasts are obtained from 1990 to 2006.}\label{fig:decamp}
\end{figure}

In Figure~\ref{fig:decamp}, we present principal components and their associated scores for the Australian fertility data from 1921 to 1989. Based on these data, the forecasts of fertility rates from 1990 to 2006 are obtained. Although the optimal $K$ is selected by the ratio-based estimator, we display only the first two components for the ease of presentation. In the top panel, we show the age profile. In the middle panel, we display the time trend of the principal component scores. In particular, the point forecasts of the scores are shown in solid line, whereas the dark and light gray regions represent the 80\% and 95\% point-wise prediction intervals, respectively. In the bottom panel, the forecasts of fertility are obtained by multiplying the fixed principal components by the forecast principal component scores before adding the main effect.

The first principal component models the fertility rates at young ages, whereas the second principal component models the fertility rates at older ages. From the forecast first principal component scores, it is clear that the fertility trend at young ages is likely to decline. Based on the forecast second principal component scores, it is evident that the fertility trend at older ages is likely to increase.

\section{Result}\label{sec:4}

\subsection{Selection of the optimal parameter}\label{sec:3.2}

Section~\ref{sec:3.1} presents one method for modeling and forecasting age-specific fertility, but the main contribution is to present a method to select optimal transformation parameter based on in-sample forecast accuracy. To investigate the in-sample forecast accuracy, we implement the rolling origin approach. Using the initial training sample in the Australian age-specific fertility, we produce one- to 17-step-ahead point and interval forecasts. Then, we increase the sample size by one year, re-estimate the model and produce one- to 16-step-ahead forecasts. This process is iterated until the training sample reaches the last year of the validation sample. This would produce 17 one-step-ahead forecasts, 16 two-step-ahead forecasts, up to one 17-step-ahead forecast. We use these forecasts to evaluate the out-of-sample forecast accuracy. For a range of forecast horizons, we calculate its forecast accuracy based on an error measure, such as MAFE or interval score given in~\eqref{eq:int_score}, over different ages and years in the validation sample. The optimal Box-Cox transformation parameter is the one that minimizes the median of a forecast error measure over a range of forecast horizons. Computationally, the optimization can be achieved by using the \verb optimize \ function in {\Rx{}}.

In Table~\ref{tab:bcfert}, we present the selected Box-Cox transformation parameter $\lambda$, based on the in-sample MAFE and interval score. For the purpose of comparison, we also consider the log transformation which is commonly used in modeling age-specific mortality. Based on the averaged MAFE and averaged interval score across 17 horizons, we found that with the selected Box-Cox transformation parameter, the out-of-sample point and interval forecast errors can be reduced in comparison with the log transformation for each forecast horizon. 

\begin{table}[ht]
\centering
\tabcolsep =0.14in
\begin{tabular}{@{}lrrrrrr@{}}
  \toprule
  & \multicolumn{3}{c}{Point forecast accuracy} & \multicolumn{2}{c}{Interval forecast accuracy} \\
 $h$  & MAFE$_{\lambda=0.46}$ & MAFE$_{\lambda=0}$ & MAFE$_{\lambda=0.4}$ & score$_{\lambda=0.46}$ &  score$_{\lambda=0}$
\\  \toprule
  1 & \textBF{0.00117} & 0.00235 & 0.00120 & \textBF{0.00543} & 0.00682 \\ 
  2 & \textBF{0.00152} & 0.00304 & 0.00155 & \textBF{0.00732} & 0.00982 \\ 
  3 & \textBF{0.00219} & 0.00388 & 0.00225 & \textBF{0.00936} & 0.01332 \\ 
  4 & \textBF{0.00285} & 0.00487 & 0.00289 & \textBF{0.01174} & 0.01696 \\ 
  5 & \textBF{0.00352} & 0.00590 & 0.00360 & \textBF{0.01412} & 0.01905 \\ 
  6 & \textBF{0.00414} & 0.00721 & 0.00432 & \textBF{0.01651} & 0.02135 \\ 
  7 & \textBF{0.00487} & 0.00853 & 0.00508 & \textBF{0.01942} & 0.02568 \\ 
  8 & \textBF{0.00564} & 0.00964 & 0.00591 & \textBF{0.02134} & 0.02909 \\ 
  9 & \textBF{0.00635} & 0.01067 & 0.00662 & \textBF{0.02400} & 0.03324 \\ 
  10 & \textBF{0.00697} & 0.01180 & 0.00735 & \textBF{0.02610} & 0.03743 \\ 
  11 & \textBF{0.00758} & 0.01291 & 0.00780 & \textBF{0.02865} & 0.04313 \\ 
  12 & \textBF{0.00819} & 0.01400 & 0.00844 & \textBF{0.03097} & 0.04745 \\ 
  13 & \textBF{0.00894} & 0.01499 & 0.00938 & \textBF{0.03314} & 0.05058 \\ 
  14 & \textBF{0.00962} & 0.01668 & 0.01013 & \textBF{0.03584} & 0.06074 \\ 
  15 & \textBF{0.01032} & 0.01782 & 0.01070 & \textBF{0.03912} & 0.06641 \\ 
  16 & \textBF{0.01068} & 0.01871 & 0.01111 & \textBF{0.04118} & 0.07244 \\ 
  17 & \textBF{0.00992} & 0.01830 & 0.01179 & \textBF{0.04337} & 0.07312 \\ \bottomrule
  Mean & \textBF{0.00614} & 0.01067 & 0.00648 & \textBF{0.02398} & 0.03686 \\ 
  Median & \textBF{0.00635} & 0.01067 & 0.00662 & \textBF{0.02400} & 0.03324 \\ 
      \bottomrule
\end{tabular}
\caption{\onehalfspacing The estimated optimal Box-Cox transformation parameters and out-of-sample point and interval forecast accuracy for different horizons. The minimum MAFE and interval score are highlighted in bold for each horizon and the summary statistics.}\label{tab:bcfert}
\end{table}

Note that Table~\ref{tab:bcfert} is consistent with the results of \citet{HB08} who found that the best point forecast accuracy (for one-step-ahead forecasts) had $\lambda=0.4$. In comparison to $\lambda=0.4$, we found that our selected $\lambda=0.46$ gives better accuracy for each horizon, but their differences in point forecast accuracy on the testing sample are marginal.

\subsection{Application to age-specific fertility}

\begin{figure}[!htbp]
\centering
\includegraphics[width=\textwidth]{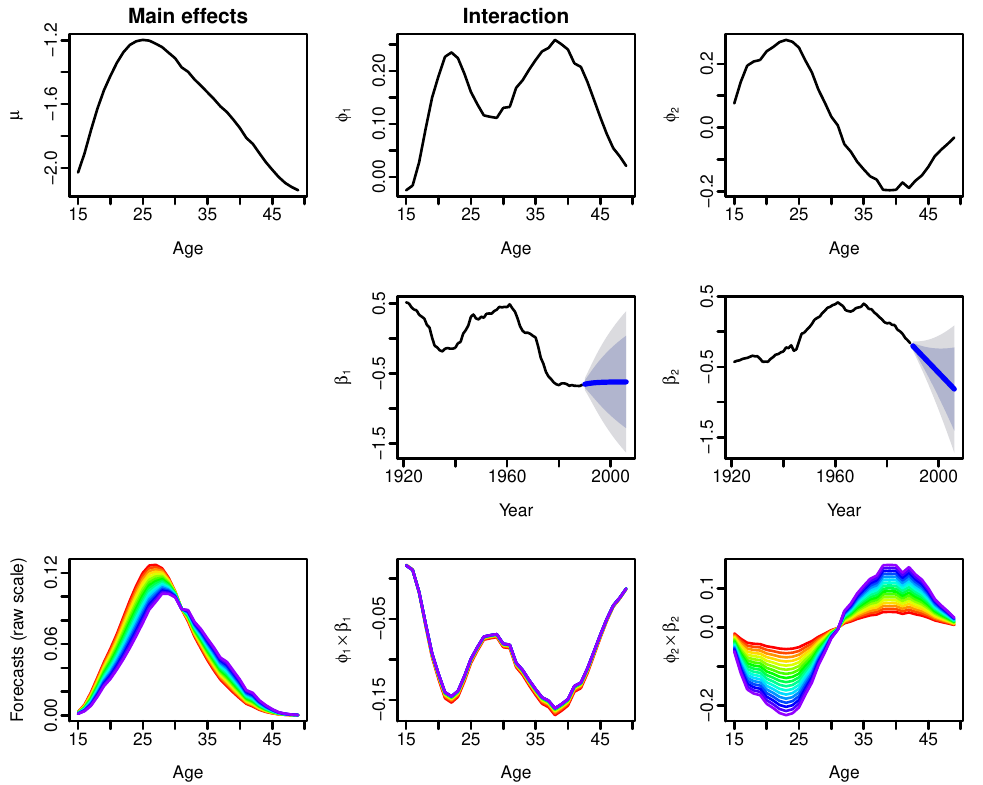}
\caption{Principal component decomposition for the transformed Australian fertility data from 1921 to 1989, from which the forecasts are obtained from 1990 to 2006.}\label{fig:decomp_BC}
\end{figure}

Prior to fitting a modified Lee-Carter model, the raw data are transformed by the Box-Cox transformation. The Box-Cox transformation may introduce a small bias, but can potentially reduce variance. As a result, this may improve estimation and forecast accuracy. In terms of its effect on forecasts of fertility,  Figure~\ref{fig:decomp_BC} displays the functional principle component decomposition for the Box-Cox transformed data with $\lambda=0.46$. From the bottom left plot, it is evident that the forecast fertility rates have a similar shape as the ones using the raw data. However, the age patterns are very different in shape from the un-transformed ones, such as the bimodality shown in the first principal component. From the forecast first principal component scores, such bimodality is likely to continue with increasing forecast uncertainties as horizon increases. The second principal component shows the contrast between ages around 25 and 40. From the forecast second principal component scores, such contrast is likely to decrease with increasing forecast uncertainties as horizon increases.

\section{Conclusion and future research}\label{sec:5}

We presented a method and an algorithm for selecting the optimal Box-Cox transformation parameter. The contributions of this paper are two-fold: First, we found that the log transformation may not be adequate for modeling and forecasting age-specific fertility. Second, we presented a way of selecting optimal Box-Cox transformation parameter based on in-sample forecast accuracy and showed that with the selected Box-Cox transformation parameter, the out-of-sample point and interval forecast accuracy can be improved. In addition, our optimal Box-Cox transformation parameter $\lambda=0.46$ produces slightly smaller point forecast error in comparison to $\lambda=0.4$ used in \cite{HB08}. 

The proposed method and algorithm can be extended to select the optimal Box-Cox transformation parameter for modeling and forecasting age-specific migration. With the selected Box-Cox transformation parameter, the forecast uncertainties associated with age-specific components of population change are more likely to be fully captured. Finally, from a Bayesian viewpoint, it is also possible to embed the selection of the optimal Box-Cox transformation parameter into the modeling and forecasting.

\section*{Acknowledgements}

The author thanks two referees for their insightful comments and suggestions, which led to a much improved manuscript. The author thanks Professor Peter W. F. Smith and Dr Jakub Bijak for insightful comments and suggestions, and Bridget Browne for proof-reading an early version of this manuscript.

\newpage

\bibliographystyle{chicago}
\bibliography{BoxCox}

\begin{thebibliography}{}

\bibitem[\protect\citeauthoryear{Bell}{Bell}{1992}]{Bell92}
Bell, W. (1992).
\newblock {ARIMA and principal components models in forecasting age-specific
  fertility}.
\newblock In N.~Keilman and H.~Cruijsen (Eds.), {\em {National Population
  Forecasting in Industrialized Countries}}, pp.\  177--200. Amsterdam: Swets
  \& Zeitlinger.

\bibitem[\protect\citeauthoryear{Booth}{Booth}{1984}]{Booth84}
Booth, H. (1984).
\newblock {Transforming Gompertz's function for fertility analysis: the
  development of a standard for the relational Gompertz function}.
\newblock {\em Population Studies\/}~{\em 38\/}(3), 495--506.

\bibitem[\protect\citeauthoryear{Booth}{Booth}{2006}]{Booth06}
Booth, H. (2006).
\newblock Demographic forecasting: 1980-2005 in review.
\newblock {\em International Journal of Forecasting\/}~{\em 22\/}(3), 547--581.

\bibitem[\protect\citeauthoryear{Box}{Box}{1988}]{Box88}
Box, G. E.~P. (1988).
\newblock Signal-to-noise ratios, performance criteria, and transformation
  (with discussion).
\newblock {\em Technometrics\/}~{\em 30\/}(1), 1--40.

\bibitem[\protect\citeauthoryear{Box and Cox}{Box and Cox}{1964}]{BC64}
Box, G. E.~P. and D.~R. Cox (1964).
\newblock An analysis of transformation.
\newblock {\em Journal of the Royal Statistical Society. Series B\/}~{\em
  26\/}(2), 211--252.

\bibitem[\protect\citeauthoryear{Bozik and Bell}{Bozik and Bell}{1987}]{BB87}
Bozik, J. and W.~Bell (1987).
\newblock Forecasting age specific fertility using principal components.
\newblock In {\em Proceedings of the American Statistical Association. Social
  Statistics Section}, San Francisco, CA, pp.\  396--401.

\bibitem[\protect\citeauthoryear{Brass}{Brass}{1981}]{Brass81}
Brass, W. (1981).
\newblock {The use of the Gompertz relational model to estimate fertility}.
\newblock In {\em International Population Conference}, Manila, pp.\  345--362.

\bibitem[\protect\citeauthoryear{Cairns, Blake, and Dowd}{Cairns
  et~al.}{2006}]{CBD06}
Cairns, A. J.~G., D.~Blake, and K.~Dowd (2006).
\newblock {A two-factor model for stochastic mortality with parameter
  uncertainty: Theory and calibration}.
\newblock {\em Journal of Risk and Insurance\/}~{\em 73\/}(4), 687--718.

\bibitem[\protect\citeauthoryear{Coale and Trussell}{Coale and
  Trussell}{1974}]{CT74}
Coale, A.~J. and T.~J. Trussell (1974).
\newblock {Model fertility schedules: Variations in the age structure of
  childbearing in human populations}.
\newblock {\em Population Index\/}~{\em 40\/}(2), 185--258.

\bibitem[\protect\citeauthoryear{Congdon}{Congdon}{1990}]{Congdon90}
Congdon, P. (1990).
\newblock {Graduation of fertility schedules: An analysis of fertility patterns
  in London in the 1980s and an application to fertility forecasts}.
\newblock {\em Regional Studies\/}~{\em 24\/}(4), 311--326.

\bibitem[\protect\citeauthoryear{Congdon}{Congdon}{1993}]{Congdon93}
Congdon, P. (1993).
\newblock Statistical graduation in local demographic analysis and projection.
\newblock {\em Journal of the Royal Statistical Society, Series A\/}~{\em
  156\/}(2), 237--270.

\bibitem[\protect\citeauthoryear{Gneiting and Katzfuss}{Gneiting and
  Katzfuss}{2014}]{GK14}
Gneiting, T. and M.~Katzfuss (2014).
\newblock Probabilistic forecasting.
\newblock {\em Annual Review of Statistics and Its Application\/}~{\em 1},
  125--151.

\bibitem[\protect\citeauthoryear{Gneiting and Raftery}{Gneiting and
  Raftery}{2007}]{GR07}
Gneiting, T. and A.~E. Raftery (2007).
\newblock {Strictly proper scoring rules, prediction, and estimation}.
\newblock {\em Journal of the American Statistical Association\/}~{\em
  102\/}(477), 359--378.

\bibitem[\protect\citeauthoryear{Hurvich and Tsai}{Hurvich and
  Tsai}{1989}]{HT89}
Hurvich, C.~M. and C.-L. Tsai (1989).
\newblock Regression and time series model selection in small samples.
\newblock {\em Biometrika\/}~{\em 76\/}(2), 297--307.

\bibitem[\protect\citeauthoryear{Hyndman and Booth}{Hyndman and
  Booth}{2008}]{HB08}
Hyndman, R.~J. and H.~Booth (2008).
\newblock Stochastic population forecasts using functional data models for
  mortality, fertility and migration.
\newblock {\em International Journal of Forecasting\/}~{\em 24\/}(3), 323--342.

\bibitem[\protect\citeauthoryear{Hyndman and Khandakar}{Hyndman and
  Khandakar}{2008}]{HK08}
Hyndman, R.~J. and Y.~Khandakar (2008).
\newblock {Automatic time series forecasting: the forecast package for R}.
\newblock {\em Journal of Statistical Software\/}~{\em 27\/}(3).

\bibitem[\protect\citeauthoryear{Hyndman and {Ullah}}{Hyndman and
  {Ullah}}{2007}]{HU07}
Hyndman, R.~J. and M.~{Ullah} (2007).
\newblock Robust forecasting of mortality and fertility rates: A functional
  data approach.
\newblock {\em Computational Statistics and Data Analysis\/}~{\em 51},
  4942--4956.

\bibitem[\protect\citeauthoryear{Keilman and Pham}{Keilman and
  Pham}{2000}]{KP00}
Keilman, N. and D.~Q. Pham (2000).
\newblock Predictive intervals for age-specific fertility.
\newblock {\em European Journal of Population\/}~{\em 16\/}(1), 41--66.

\bibitem[\protect\citeauthoryear{Knudsen, McNown, and Rogers}{Knudsen
  et~al.}{1993}]{KMR93}
Knudsen, C., R.~McNown, and A.~Rogers (1993).
\newblock {Forecasting fertility: An application of time series methods for
  parameterized model schedules}.
\newblock {\em Social Science Research\/}~{\em 22\/}(1), 1--23.

\bibitem[\protect\citeauthoryear{Lam, Yao, and Bathia}{Lam
  et~al.}{2011}]{LYB11}
Lam, C., Q.~Yao, and N.~Bathia (2011).
\newblock Estimation of latent factors in high-dimensional time series.
\newblock {\em Biometrika\/}~{\em 98\/}(4), 901--918.

\bibitem[\protect\citeauthoryear{Lee}{Lee}{1993}]{Lee93}
Lee, R.~D. (1993).
\newblock {Modeling and forecasting the time series of US fertility: Age
  distribution, range and ultimate level}.
\newblock {\em International Journal of Forecasting\/}~{\em 9\/}(2), 187--202.

\bibitem[\protect\citeauthoryear{Lee and Carter}{Lee and Carter}{1992}]{LC92}
Lee, R.~D. and L.~R. Carter (1992).
\newblock {Modeling and forecasting U.S. mortality}.
\newblock {\em Journal of the American Statistical Association\/}~{\em
  87\/}(419), 659--671.

\bibitem[\protect\citeauthoryear{Murphy}{Murphy}{1982}]{Murphy82}
Murphy, M.~J. (1982).
\newblock {Gompertz and Gompertz relational models for forecasting fertility:
  An empirical exploration}.
\newblock Working paper, {Centre for Population Studies, London School of
  Hygiene and Tropical Medicine}, London.

\bibitem[\protect\citeauthoryear{{R Core Team}}{{R Core Team}}{2014}]{Team13}
{R Core Team} (2014).
\newblock {\em R: A Language and Environment for Statistical Computing}.
\newblock Vienna, Austria: R Foundation for Statistical Computing.
\newblock \url{http://www.R-project.org/}.

\bibitem[\protect\citeauthoryear{Shang, Booth, and Hyndman}{Shang
  et~al.}{2011}]{SBH11}
Shang, H.~L., H.~Booth, and R.~J. Hyndman (2011).
\newblock {Point and interval forecasts of mortality rates and life expectancy:
  A comparison of ten principal component methods}.
\newblock {\em Demographic Research\/}~{\em 25}, 173--214.

\bibitem[\protect\citeauthoryear{Shang and Hyndman}{Shang and
  Hyndman}{2013}]{SH13}
Shang, H.~L. and R.~J. Hyndman (2013).
\newblock {\em rainbow: Rainbow plots, bagplots and boxplots for functional
  data}.
\newblock R package version 3.2.
  \url{http://cran.r-project.org/web/packages/rainbow}.

\bibitem[\protect\citeauthoryear{Thompson, Bell, Long, and Miller}{Thompson
  et~al.}{1989}]{TBL+89}
Thompson, P.~A., W.~R. Bell, J.~F. Long, and R.~B. Miller (1989).
\newblock Multivariate time series projections of parameterized age-specific
  fertility rates.
\newblock {\em Journal of the American Statistical Association\/}~{\em
  84\/}(407), 689--699.

\bibitem[\protect\citeauthoryear{Zeng, Wang, Ma, and Chen}{Zeng
  et~al.}{2000}]{ZWM+00}
Zeng, Y., Z.~Wang, Z.~Ma, and C.~Chen (2000).
\newblock {A simple method for projecting or estimating $\alpha$ and $\beta$:
  An extension of the Brass relational Gompertz fertility model}.
\newblock {\em Population Research and Policy Review\/}~{\em 19\/}(6),
  525--549.

\end{thebibliography}

\end{document}